\documentclass[summary]{URSIGASS2020}

\usepackage{url}
\usepackage{pdfpages}
\usepackage{grffile}
\usepackage{amsmath,amssymb}
\usepackage{amsfonts}
\usepackage{bm}
\usepackage{bbm}
\usepackage[ruled,linesnumbered]{algorithm2e}
\usepackage{bbold}
\usepackage{url}
 \usepackage{arydshln}

\newdimen\Lwidth
\newdimen\Lwidthtwo
\newdimen\Lwidththree
\newcommand{\vect}[1]{\boldsymbol{#1}}

\def\argmin{\mathop{\rm argmin}}

\title{Offline and online reconstruction for radio interferometric imaging}

\author{Xiaohao Cai\affref{ref1}, Luke Pratley\affref{ref2},
  and  Jason D. McEwen\affref{ref1}}

\affiliation{%
  \aff{ref1}{Mullard Space Science Laboratory (MSSL),  University College London, United Kingdom}
  \aff{ref2}{Dunlap Institute for Astronomy and Astrophysics, University of Toronto, Canada}
}

\begin{document}

\maketitle

\begin{abstract}
	Radio astronomy is transitioning to a big-data era due to the emerging generation of radio interferometric (RI) telescopes, such as the 
	Square Kilometre 	Array (SKA), which will acquire massive volumes of data. In this article we review methods proposed recently to resolve the ill-posed inverse problem of imaging the raw visibilities acquired by 
	RI telescopes in the big-data scenario. We focus on the recently proposed online reconstruction method \cite{CPM19} and the considerable savings in data storage requirements and computational cost that it yields.
\end{abstract}

\section{Introduction}\label{sec:intro}
Radio astronomy has transitioned from the first observations in the 1930s to a data-rich era,  and is 
transitioning to the so-called big-data era in coming years, due to rapid technological developments.
For example, representative next-generation radio interferometric (RI) telescopes --
the LOw Frequency ARray (LO-FAR\footnote{\url{http://www.lofar.org}}, \cite{van13}),
the Extended Very Large Array (EVLA\footnote{\url{http://www.aoc.nrao.edu/evla}}),
the Australian Square Kilometre Array Pathfinder (ASKAP\footnote{\url{http://www.atnf.csiro.au/projects/askap}}), 
the Murchison Widefield Array (MWA\footnote{\url{http://www.mwatelescope.org/telescope}}), 
and the Square Kilometer Array (SKA\footnote{\url{http://www.skatelescope.org}}) -- 
will acquire large volumes of data, and achieve significantly higher dynamic range and angular resolution 
than previous generations. This new generation of radio telescopes will bring further challenges in imaging and scientific analysis.

Radio interferometers, briefly speaking, sample Fourier coefficients (visibilities) of the radio brightness distribution in the sky. 
It is then necessary to solve an ill-posed linear inverse problem to reconstruct the image of the sky from acquired visibilities  \cite{tho08}.
In the era of big-data, the enormous data rates will create practical challenges in both data storage requirements and computational cost. 

In this summary we review several methods, particularly the online reconstruction method \cite{CPM19}, proposed recently to address 
the ill-posed inverse problem in RI imaging in the big-data scenario.  We pay particular attention to data storage requirements and computational cost.
	
\section{RI imaging and offline methods}\label{sec:ri}
A fundamental problem in RI imaging is to recover an image (sky brightness), $\vect x \in \mathbb{R}^N$, 
from the visibilities, $\vect y \in \mathbb{C}^{M}$, measured by telescopes, which  
raises an ill-posed inverse problem with the following representation,
\begin{equation}\label{eqn:y}
	{\vect y}=\bm{\mathsf{\Phi}} {\vect x} + {\vect n},
\end{equation}
where $\bm{\mathsf{\Phi}} \in \mathbb{C}^{M\times N}$ models the telescope measurement process and ${\vect n} \in \mathbb{C}^{M}$ represents additive noise. 
Without loss of generality, we split the measurements ${\vect y}$ into $B$ blocks and assume these blocks are received 
at different but consecutive time slots, {\it i.e.},
\begin{align} \label{eqn:split-y}
	{\vect y} & = \begin{bmatrix} 
	{\vect y}_1^{\top}, \cdots, {\vect y}_{k}^{\top}, \cdots,
	{\vect y}_{B}^{\top}
	\end{bmatrix}^{\top}, 
	\ \ {\vect y}_{k} \in \mathbb{C}^{M_k},
\end{align}
where ${\vect y}_{k}$ is received earlier than ${\vect y}_{k+1}$ and $\sum_{k=1}^B M_k = M$.

The sparsity property of $\vect x$ under 
a basis or dictionary, $\bm{\mathsf{\bm{\mathsf{\Psi}}}} \in \mathbb{C}^{N\times L}$, is an effective prior \cite{PMD18} 
to consider when solving the 
inverse problem, {\it i.e.}, ${\vect x} = \bm{\mathsf{\Psi}} {\vect a} = \sum_{i} \bm{\mathsf{\Psi}}_i a_i$, where 
vector ${\vect a} = (a_1, \cdots, a_L)^\top$ represents the synthesis coefficients
of ${\vect x}$ under $\bm{\mathsf{\Psi}}$ with the sparsity prior that many coefficients of $\vect a$ are nearly zero. 
For further details regarding RI imaging see, {\it e.g.}, \cite{tho08} and references therein. 

A number of models have been proposed to recover the underlying image $\vect x$. For example, 
the underlying image $\vect x$ can be recovered by solving the following constrained model
\begin{equation}\label{eqn:ir-con-af}
	{\vect x}^* = \argmin_{\vect x} \|\bm{\mathsf{\Psi}}^\dagger {\vect x}\|_1,  
		\quad {\rm s.t.} \ \  \|{\vect y}-\bm{\mathsf{\Phi}} {\vect x}\|_2^2/2\sigma^2 \le \epsilon,
\end{equation}
where $\|\cdot \|_1$ is the $\ell_1$-norm promoting sparseness and $\sigma$ represents the standard deviation of the noise.
The underlying image $\vect x$ can also be recovered using the {\it maximum-a-posteriori} (MAP) estimation, {\it i.e.}, 
solving the unconstrained model
\begin{equation}\label{eqn:ir-un-af}
	{\vect x}^* = \argmin_{\vect x} \Big\{\mu \|\bm{\mathsf{\Psi}}^\dagger {\vect x}\|_1 
	+ \|{\vect y}-\bm{\mathsf{\Phi}} {\vect x}\|_2^2/2\sigma^2 \Big\},
\end{equation}
where $\mu$ is the regularisation parameter used to balance the tradeoff between sparsity and data fidelity, 
see {\it e.g.}, \cite{CPM18b} and references therein. 

\subsection{Offline methods}
\begin{figure*}
  \begin{center}
    \begin{tabular}{cc}
      	 \includegraphics[trim={{.55\Lwidth} {1.0\Lwidth} {1.2\Lwidth} {.86\Lwidth}}, clip, width=0.6\Lwidth, height = 0.9\Lwidth]
		{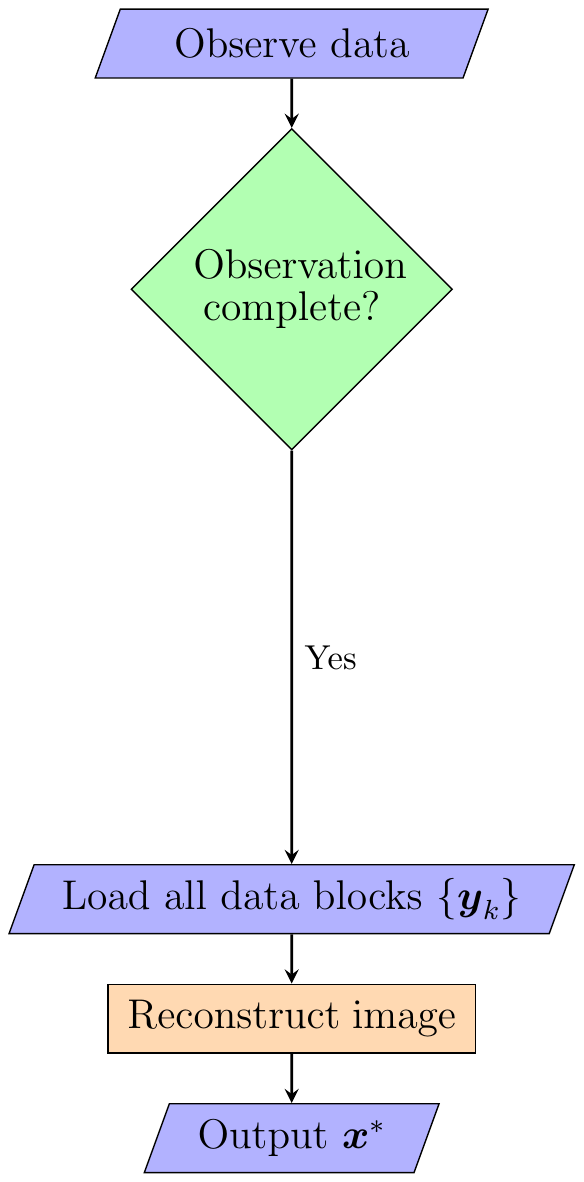} \hspace{0.5in} &
  	 \includegraphics[trim={{.55\Lwidth} {1.03\Lwidth} {.95\Lwidth} {.87\Lwidth}}, clip, width=0.8\Lwidth, height = 0.9\Lwidth]
		{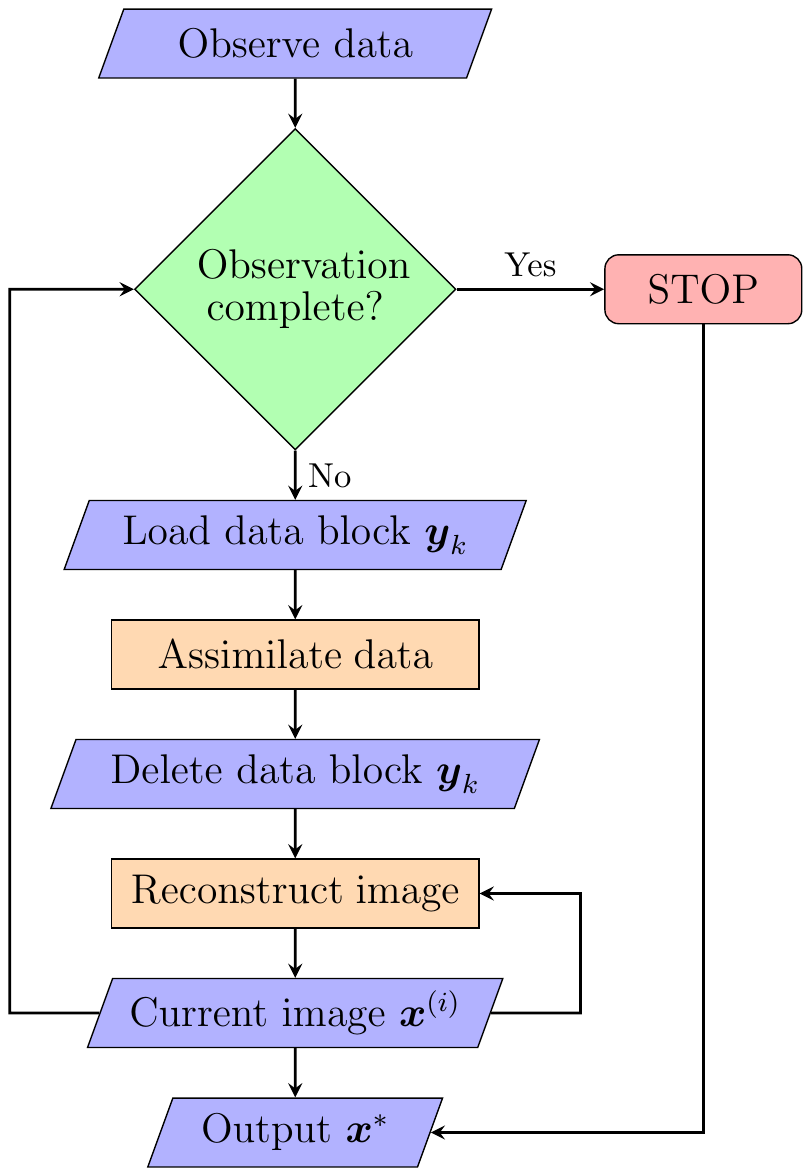}  \\
		(a) Offline method \hspace{0.5in}  & (b) Online method (credit \cite{CPM19})
    \end{tabular}
  \end{center}
  \caption{Methods for RI imaging. Panels (a) and (b) show offline and online imaging methods, respectively.}
  \label{fig:offline_online_diag}
\end{figure*}

Classical image reconstruction methods, such as CLEAN-based methods \cite{hog74,BC04} and the
maximum entropy method (MEM) \cite{A74}, have served the community well but do not exploit modern image reconstruction techniques and 
struggle to confront the upcoming big-data era.
Recently, compressive sensing (CS) techniques have been considered for RI imaging, exploiting sparse regularisation techniques,
and have shown promising results \cite{PMD18} compared to traditional approaches such as CLEAN.
Typically they make use of constrained models like (\ref{eqn:ir-con-af}) and were developed to scale to big-data \cite{PMD19}, as anticipated from the SKA, using, 
{\it e.g.}, distribution, parallelisation, dimensionality reduction, and/or stochastic strategies.  
In \cite{CPM18,CPM18b}, Bayesian inference techniques for sparsity-promoting priors were presented to quantify the 
uncertainties associated with reconstructed images, {\it e.g.} to estimate local credible intervals ({\it cf.} error bars) on
recovered pixels, where non-constraint models like (\ref{eqn:ir-un-af}) are adopted. In particular, in \cite{CPM18b}, 
MAP estimation techniques were presented to scale uncertainty quantification to 
massive data sizes, {\it i.e.} to big-data. 

All of these reconstruction methods ({\it e.g.}, CLEAN, MEM and CS-based methods) can be categorised as {\it offline} methods.
They need to store the entire set of observed visibilities for subsequent processing once data acquisition is completed 
({\it i.e.}, after the full observation is made, often $\sim$10 hours or longer), as shown in Figure \ref{fig:offline_online_diag} (a). 
In other words, data storage is highly demanding for offline methods even if 
state-of-the-art techniques such as distribution and parallelisation are exploited. 

{\it Online} methods, processing data piece-by-piece as they are acquired without having the entire data-set available from 
the start, have great potential for RI imaging.

\section{Online reconstruction for RI imaging}\label{sec:online}
Online reconstruction \cite{CPM19} has great potential for RI imaging, owing to its natural ability to manage two main issues 
in RI imaging: (1) the time of acquiring the measurements $\vect y$ can be long (often $\sim$10 hours or longer),
and (2) the space needed to store the data can be extremely large, particularly in the big-data era. 
Offline methods wait to obtain and then store all measurements, thus requiring significant storage and computational cost that cannot be avoided.

\subsection{Online methodology}
The online methodology for RI imaging is shown in the diagram in Figure \ref{fig:offline_online_diag} (b). 
As is shown, firstly, the algorithm checks whether the data observation stage has completed. If yes, no new data block will be 
observed and thus the online method stops. Otherwise, the algorithm: loads the new observed data block;
assimilates it; releases the data block; 
updates the intermediate reconstructed image (using the newly assimilated data); 
and then sets the current reconstructed image as the starting point for the next iteration. 
The above steps are repeated until the data observation stage completes and then the final reconstructed image is set as the output.

Clearly, the online method starts at the beginning of the data observation stage not the end; therefore the waiting time that the 
offline methods encounter is utilised by the online method for computation. Moreover, after assimilating each data block in the online method, 
that data block is allowed to be released (deleted) immediately, which resolves the difficulty of storing the whole data-set that offline methods struggle with in the big-data era. 

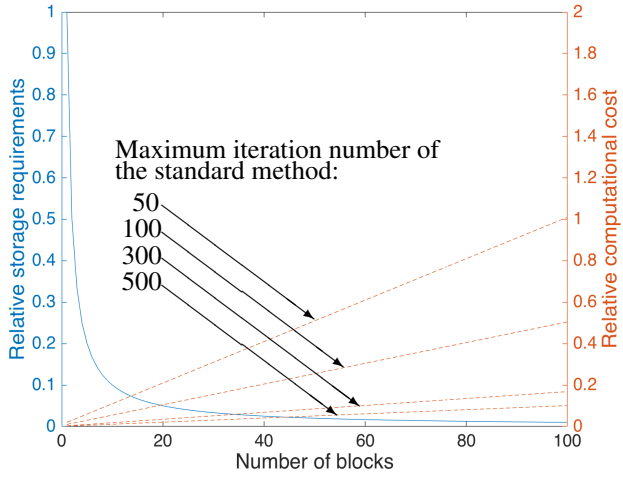
\begin{figure}[htbp]
  \centering
		\includegraphics[trim={{.055\Lwidthtwo} {.02\Lwidthtwo} {.02\Lwidthtwo} {.05\Lwidthtwo}}, clip, width=0.9\Lwidthtwo, height = 0.70\Lwidthtwo]
		{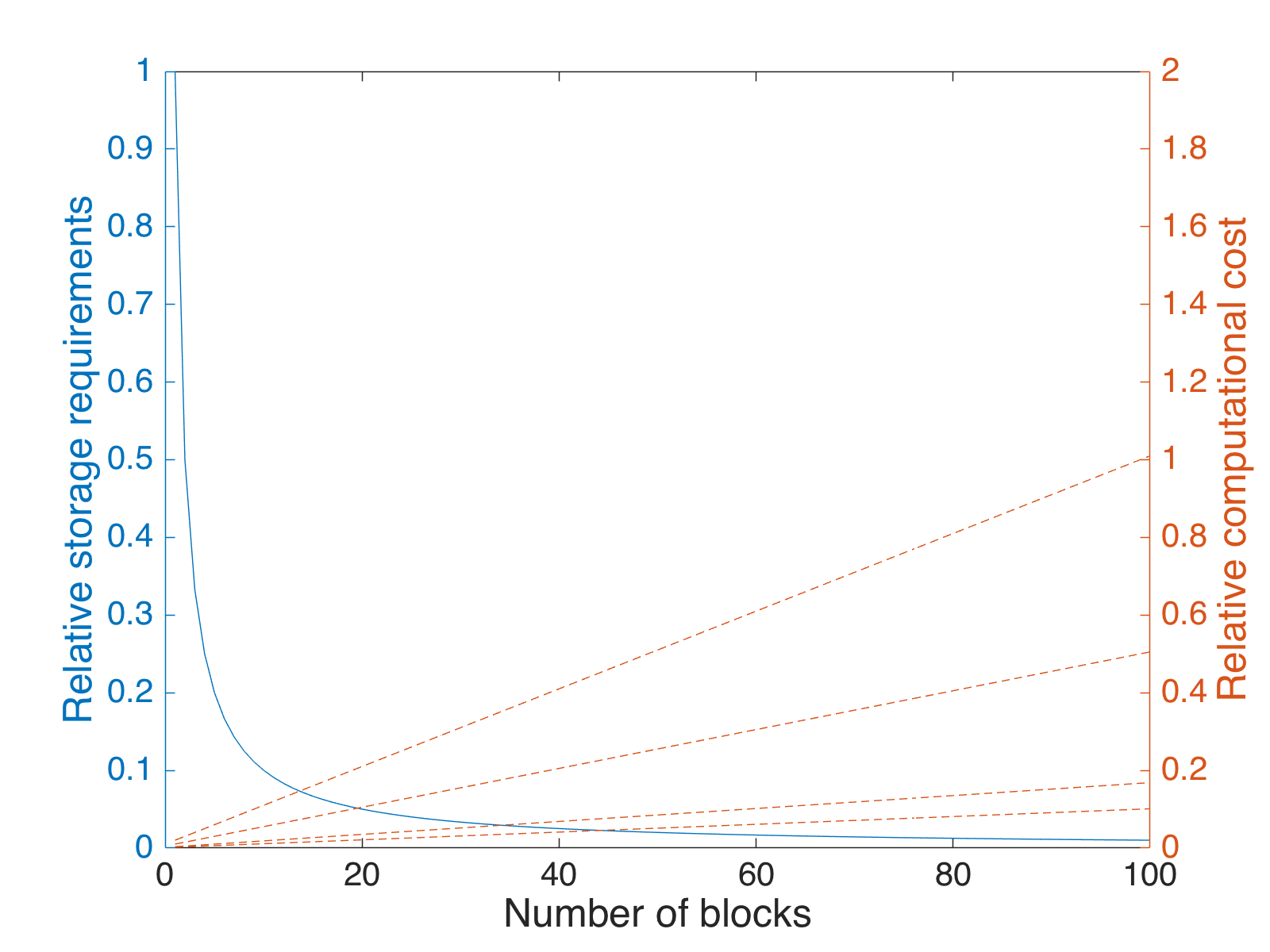} 
		\put(-190,120){Maximum iteration number of} 
		\put(-190,112){the standard method:} 
		\put(-186,100){ 50}  \put(-175,102){ \vector(4,-3){57}} 
		\put(-190,90){ 100}  \put(-175,92){ \vector(4,-3){68}} 
		\put(-190,80){ 300} \put(-175,82){ \vector(4,-3){74}} 
		\put(-190,70){ 500} \put(-175,72){ \vector(4,-3){65.5}}
  \caption{Comparison between the offline method and the online method in terms of visibility storage requirements
		and computational cost (credit \cite{CPM19}). In the plot, the left vertical-axis represents the ratio of visibility storage requirements 
		between the online method with different number of visibility blocks and the offline methods (blue solid curve); 
		the right vertical-axis represents the approximate ratio of computational cost between the online method and the offline 
		methods with different maximum iteration numbers (brown dashed lines).}
		\label{fig-ite-comp-memory}
\end{figure}

\subsection{Data storage requirements}
The above discussion tells us that the offline methods need to store the whole data, while the online method 
just needs space for the current data block. In essence, if all blocks are the same size, the storage requirement 
for the online algorithm is $1/B$ of the total number of visibilities (recall $B$ is the number of visibility blocks).
Figure \ref{fig-ite-comp-memory} (the blue solid curve) shows the ratio of visibility storage requirements 
between the online method and the offline methods for different number of visibility blocks. 

Other important advantages of the online method in terms of storage requirement are that:
(1) it has the potential of tackling RI imaging problems encountered with an arbitrarily large amount of visibilities;
and (2) it is able to immediately process any new observed visibilities. On the contrary, offline methods cannot address these issues.

\subsection{Computational cost}
Comparing to offline methods, the online method can also provide considerable computational savings when the 
number of visibility blocks considered is not much larger than the number of iterations necessary for the offline methods 
since the amount of data to be considered for early iterations is small.
See Figure~\ref{fig-ite-comp-memory} (the brown dashed lines) for the pictorial explanation of the comparison between the online 
and offline methods in terms computational cost.

It is worth mentioning that the online method actually has the potential to achieve a reconstruction as soon as 
no more visibility blocks are available ({\it i.e.}, immediately once the observation is complete), since the online method executes almost all of its iterations before the data acquisition stage finishes. On the contrary, all of the computational costs of the offline methods have to be carried out after the data acquisition stage.

\subsection{Reconstruction quality}
\begin{figure}[htbp]
  \centering
  \begin{tabular}{cc}
  		\includegraphics[trim={{.15\Lwidththree} {.07\Lwidththree} {.02\Lwidththree} {.072\Lwidththree}}, clip, width=0.45\Lwidththree, height = 0.40\Lwidththree]
		{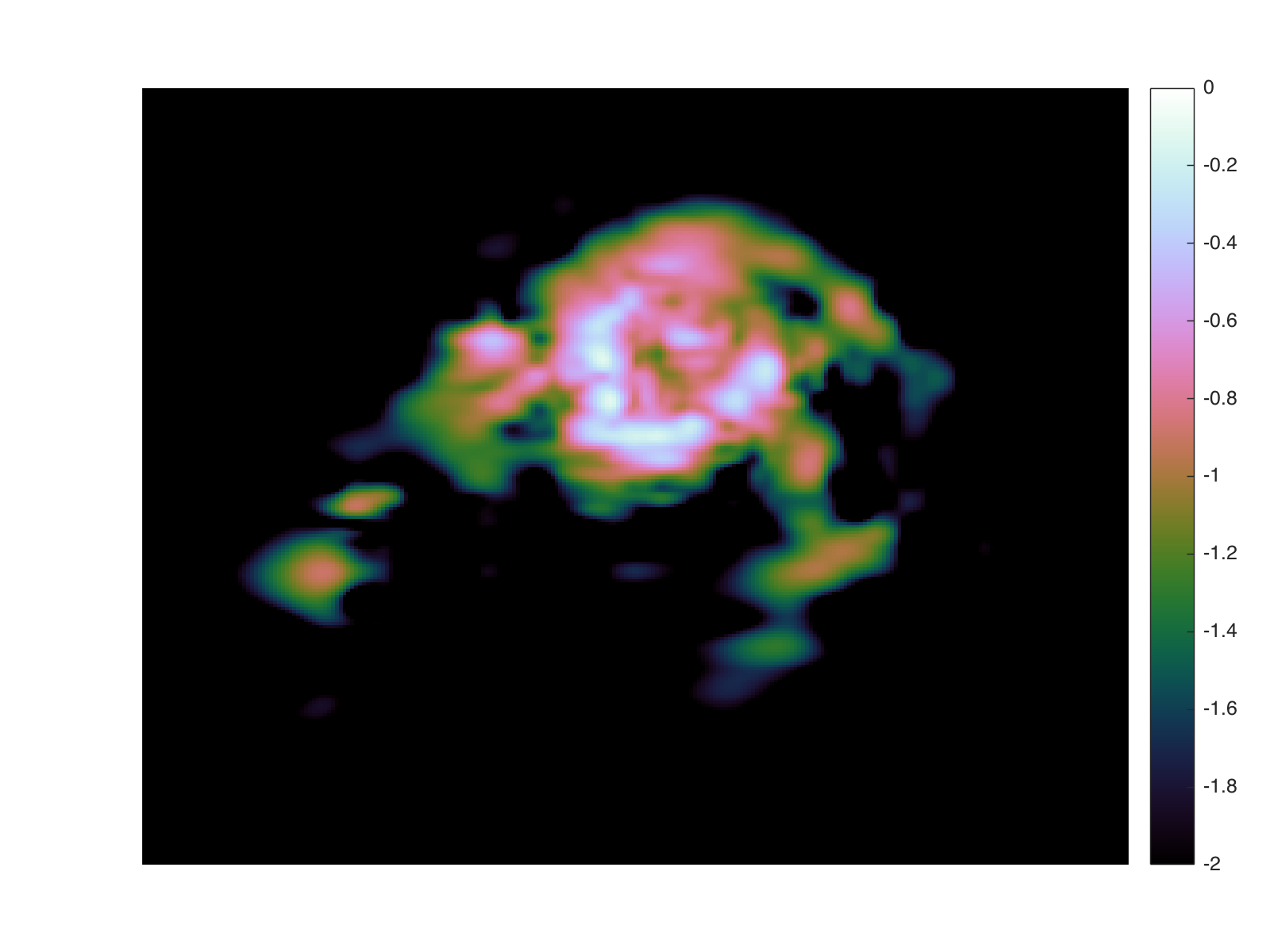} &
		\includegraphics[trim={{.15\Lwidththree} {.07\Lwidththree} {.02\Lwidththree} {.072\Lwidththree}}, clip, width=0.45\Lwidththree, height = 0.40\Lwidththree]
		{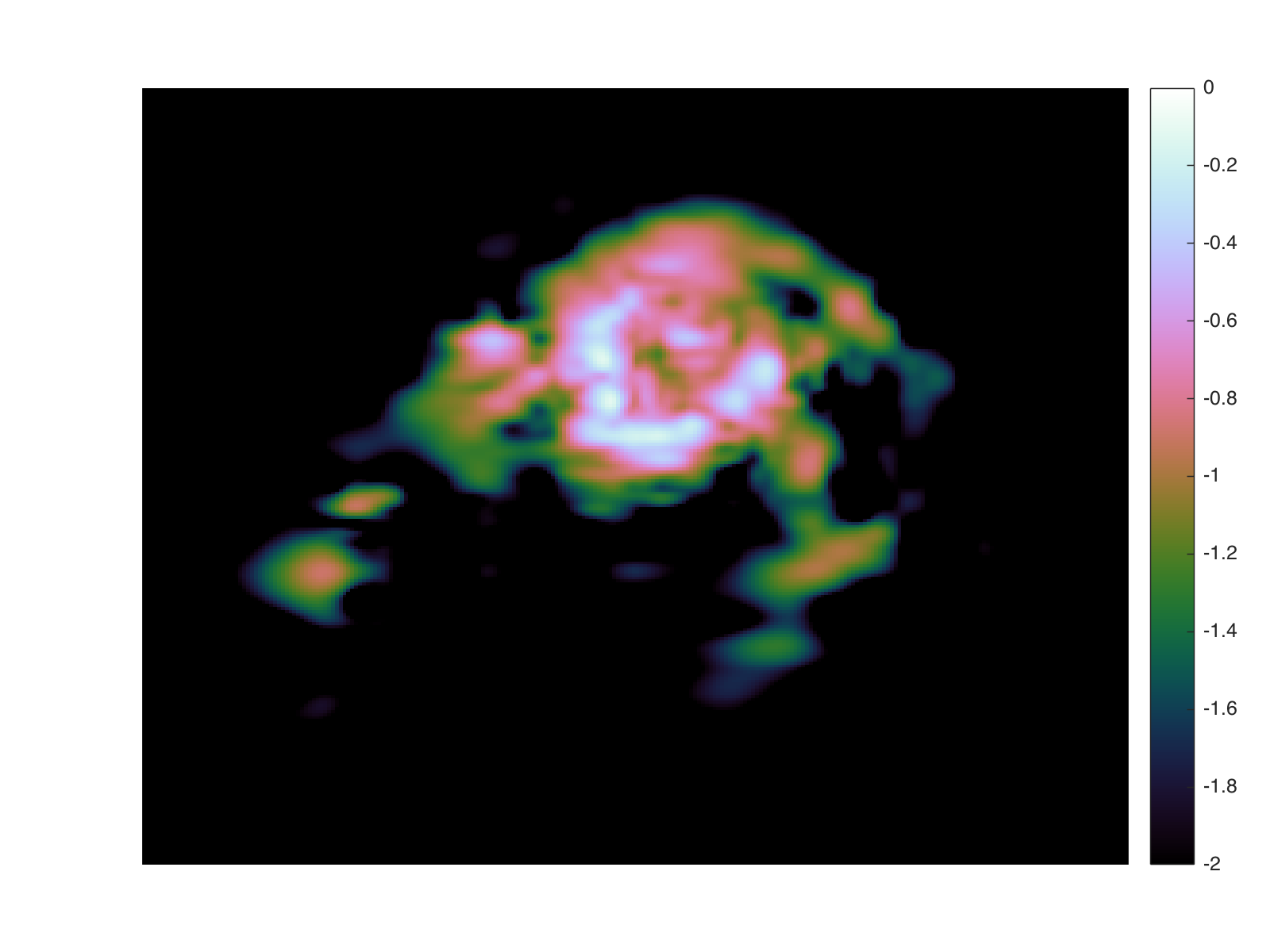}  \\
		{ (a) Offline method} & {(b) Online method}                    \\
		(storage: 100\% visibilities)   & { (storage: 2\% visibilities)} 
  
  \end{tabular}
  \caption{Image reconstruction results of the offline method and the online method
		for image M31 (credit \cite{CPM19}). The unconstrained model \eqref{eqn:ir-un-af} is adopted. The number of iterations for the
		tested methods is set to 50. All images are shown in ${\tt log}_{10}$ scale. 
		Panel (a): result of the offline method, requiring storage for 100\%  of acquired visibilities at once.
		Panel (b): result of the online method, with visibilities gradually increased from 2\% to 100\% block-by-block, requiring storage of 2\% of all acquired visibilities. }
		\label{fig-m31}
\end{figure}

Theoretically, the images reconstructed by the online method are of the same fidelity as those recovered by the equivalent offline
methods and, in practice, very similar reconstruction fidelity is achieved.
The online method can provide very good reconstructions after processing the last visibility block, often typically as good as the quality achieved by
offline methods (see Figure \ref{fig-m31}). 
If a few additional iterations are executed for the online method the quality of the reconstruction can be further improved.
However, the improvement is not dramatic and the standard number of iterations, basically, can ensure excellent reconstructions already 
(see \cite{CPM19} for more details).

\section{Conclusions and future work}\label{sec:con}
In this article we briefly reviewed reconstruction methods -- online and offline -- for RI imaging, motivated by critical 
computational problems in scaling RI imaging to the big-data era of radio astronomy that will be ushered in 
by the SKA and precursor telescopes, in terms of storage requirements and computational cost.  

Current RI imaging methods, such as CLEAN, its variants, and compressive sensing approaches ({\it i.e.}, sparse regularisation),
have yielded excellent reconstruction fidelity. However, scaling these methods to big-data remains difficult if not impossible 
in some cases. All state-of-the-art offline methods in RI imaging lack the ability to process data streams as they are acquired during 
the data observation stage. 

The online method reviewed \cite{CPM19} starts the reconstruction task at the beginning of the data acquisition stage (not after) 
and keeps updating the quality of the reconstruction by continually assimilating newly acquired visibilities (visibility blocks), 
before discarding them to release storage. In other words, it combines the data acquisition stage with the data processing stage.

The online method has the advantage of significantly lower visibility storage requirements. 
In principle, the storage needed for the online method can be arbitrarily small; recall that the offline methods
always require all the visibilities to be stored for subsequent processing.
The online method also achieves good reconstruction fidelity much faster than offline methods, 
which do not begin until the visibility acquisition stage is completed. 
Roughly speaking, the online method has the ability to provide
an excellent reconstruction as soon as the visibility acquisition procedure completes.
Moreover, the computational cost of the online method is further reduced for a reasonable choice of number of blocks
since the amount of data to be considered for early iterations is small.

Consequently, these two main virtues -- extremely low storage requirements and low computational cost -- 
make the online method highly relevant for addressing the big-data processing obstacles of RI imaging in the near future. 
We anticipate online imaging techniques will be critical in scaling RI imaging to the emerging 
big-data era of radio astronomy.  

There are a number of avenues of future work. Since the proposed online framework is very general, 
it will be interesting to investigate equipping other methods with this online strategy.
{The online method will be implemented in the existing PURIFY\footnote{\url{https://github.com/astro-informatics/purify}} 
package for RI imaging, where it may then be applied easily to real observations and combined with existing performance 
gains from distributed and shared parallelisation. }
Finally, it is worth integrating the online method with the uncertainty quantification framework presented in \cite{CPM18b} to 
perform efficient imaging and uncertainty quantification for the emerging big-data era of radio astronomy.

\section{Acknowledgements}
This work is supported by the UK Engineering and Physical Sciences Research Council (EPSRC) by grant EP/M011089/1 and the Leverhulme Trust.


\end{document}